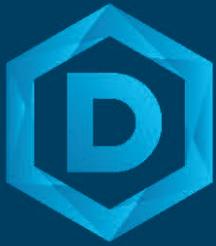

# GOLDEN EYE

## Theory of Havana Syndrome

Adam Dorian Wong
adam.wong[at]trojans.dsu.edu
(@MalwareMorghulis)

v1.2

Date: 22 FEB 2024

24-P-0558

# Dedication

I write this whitepaper

with the hopes of one day discovering the truth,

and to support the injured & nameless American patriots whom serve in silence,

not from deep hatred for our enemies,

but out of profound love of country.

# Abstract

On the global stage, the US is a superpower. It has significant influence diplomatically, economically, and militarily. American strength also comes from building partnerships and relationships through diplomacy. However, what if that hegemonic rule were to be challenged? Not necessarily militarily, but challenged covertly, asymmetrically. In 2016, the US engaged with Cuba to thaw relations that were frozen since the Cold War. However, during this push for détente, State Department diplomats in Havana, Cuba complained of suffering from a strange array of symptoms: ruptured eardrums, vertigo, dizziness, disorientation, nausea, pain, and hearing noises. This collection of symptoms was dubbed "*Havana Syndrome*". Eventually, the afflictions spread to various capital cities around the world with American presence. This research whitepaper dives into the various hypotheses of Havana Syndrome: advanced sonic weapons, direct-energy attacks, mass hysteria, crickets, or chemical exposure. Based on relevant publicly available information, open-source research, and second-hand analysis of published journals for supposed injuries, it is a personal suspicion that Havana Syndrome could be the result of a direct-energy weapon aimed at the vestibule system of the inner ear.

# Introduction

Nations will always reside within the spectrum of conflict and influenced through soft-power (diplomacy, influence, economics) and hard-power (military might). The power of the US derives from its ability to project diplomacy and build relations with partner nations and resolve differences with former enemies. In 2016, US diplomats stationed in Havana, Cuba, reported injuries outlining mysterious symptoms named: "*Havana Syndrome*". Ailments included: hearing horrid noises, vertigo, nausea, disorientation were commonly reported. Medical professionals noted striking similarities to concussion-like injuries and anomalous changes in brain white matter of victims. The incidents were not isolated to Havana. Over the course of several years, these symptoms not only affected diplomats in Cuba but across multiple countries, especially within national capital cities. Even allied diplomats from Canada reported suffering similar issues. Havana Syndrome has been an anomaly among the scientific and medical community. Many hypotheses exist on the root-cause of the problem. Some suggest it is the result of advanced sonic or direct energy weapons. Others claim it is mass hysteria and psychological distress from overseas hardship. Some believe it is the result of insects or exposure to chemicals. The current assessment is that a direct-energy or sonic weapon were unlikely. Ultimately, this whitepaper re-examines the various aspects of these hypotheses in conjunction with available medical information and reports on Havana Syndrome.

# Havana Syndrome

In 2016, staff members stationed at the US Embassy in Havana, Cuba had reported experiencing unusual symptoms ranging from nausea, vertigo, cognitive dysfunction, and ringing in the ears. Numerous US Department of State employees and medical professionals were confounded by these reports. This collective of symptoms was dubbed "Havana Syndrome". Victims reported hearing cricket-like noises, hearing screeching, hearing marbles dropping, tinnitus, disorientation, and headaches [1]. The cause is still a subject of debate [2].

| Mainstream Hypotheses | Nature of Effect |
|---|---|
| Ultra Sonic Weapon | Mechanical [3] |
| Direct-Energy (Microwave) | Electromagnetic [3] |
| Mass Hysteria | Psychological |
| Crickets | Mechanical [3] |
| Compound / Agent Exposure | Chemical |

Lives have been changed from debilitating injuries [4] [5]. Some travelers had visited numerous doctors and had symptoms that mimicked what would typically be found with concussions [6]. University of Pennsylvania evaluated the medical data from a majority of State Department victims [7] [8] [9]. There were subtle differences in brain scans, white matter, and connective tissue between regions of the brain [9] [10] [11].

Some report hearing high-pitched sounds in specific rooms of their domiciles [12]. Diplomats were "*medically confirmed*" to have been affected by ailments [13] [14].

**Current Assessment**. The current assessment, based on uniform consensus between several government organizations and research centers, suggests that foreign sonic or electromagnetic weaponry are *very unlikely* to have caused cases of Havana Syndrome [15]. A declassified 2018 report by JASON team (MITRE Corporation) was unable to conclusively identify or detect direct energy weaponry or similar artifact [16]. The assessments appear inconclusive, at best.

## H1: Ultra Sonic Weapons (USW)

**Applications in Warfare**. Although, some symptoms align to Havana Syndrome, it does not fully explain certain injuries incurred by USG officials. Sound has been utilized in warfare since antiquity such as with the biblical story of the campaigns of Gideon or Battle of Jericho [17] [18]. In WWII, the German Luftwaffe modified Stukas to sound sirens during dive bombings [18]. During the Vietnam War, sound was implemented within the jungle environment to scare oppositional force during Operation Wandering Souls [19] [20]. In 2005, Israeli Defense Forces utilized vehicle-mounted sonic weapons against Palestinians and adversarial settlers. Protesters reported having suffered symptoms of dizziness and nausea from audible pulses fired in 10 second intervals which still was heard after covering ears [21]. In recent history, the US military leveraged Metallica against adversaries in both Iraq and Afghanistan campaigns [22] [23]. Ultimately, sound provides PSYOPS value to the battlespace.

**Modern Domestic Security**. Law enforcement has used acoustic weapons as the *less-lethal* alternative for crowd-dispersal. The Long-Range Acoustic Device (LRAD) was originally developed to counter small watercraft following the tragic bombing of the USS Cole in Yemen [24] [25]. Companies like Genasys develop multiple product variations of LRAD [26]. LRADs are controversial due to lack of data or research in long-term health effects of these weapons platforms and poor discriminate use-of-force based on angular area-of-effect [27]. Additionally, ultrasonic frequencies have long been suspected of causing illness and being occupational hazards [28]. Radio Liberty places intensity of LRADs along a spectrum: 60 dB (conversation), 90 dB (lawnmower), 130 dB (pain threshold), 162 dB (max LRAD output) [24]. LRADs have been known to operate on the audible frequency between 2,000 and 4,000 Hz that causes optimal discomfort in humans [25] [29]. These devices have been known to be most painful at close distance (within 20 meters), while unintelligible at ranges as far as approximately 9000 meters [30]. Sound is mechanical and broken into three or four bands relative to the human hearing range: infrasonic (<20 Hz), audible (20 Hz to 2 kHz), audible-sensitive (2 kHz to 20 kHz), and ultrasonic (>20 kHz) [29]. In 2018, researchers in the UK studied adverse effects from exposure within the ultrasonic spectrum were mixed. There was higher general discomfort, but the study was unable to reproduce the severe publicly reported effects of nausea, disorientation, or headaches [31] [32]. Albeit, the study used a frequency range far higher than LRAD and far lower decibel level than LRAD.

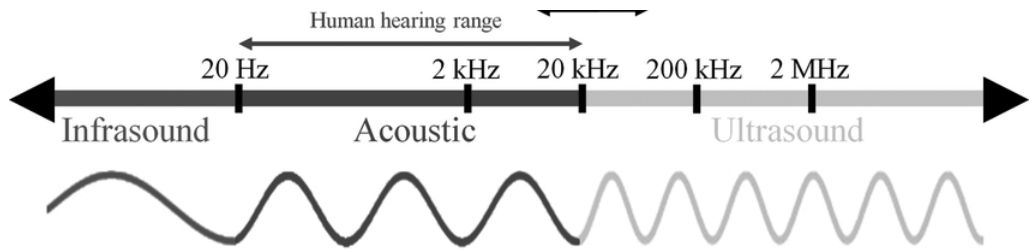
*Figure 1: Frequency of Sound [29]*

***Discussion***. The symptoms popularly reported by victims of acoustic weapons are: nausea, headaches, dizziness, and disorientation. Ultrasonic noise has been observed in killing mice at close-range. Although, symptoms align to Havana Syndrome, a problem still remains: victims continued to suffer brain damage even after leaving the perceived and affected Area of Operations. LRADs are fairly large equipment to operate. These would be detectable either with high power consumption, audio sensor triangulation, or through realization for indiscriminate targeting of adjacent unintended victims. Sound is a mechanical wave and prone to bouncing off of walls. Considering Occam's Razer, this weapon would need to be cleverly hidden from view, but obscuring the ultrasonic weapon would inhibit the effectiveness of it and having to penetrate walls. Therefore, ultrasonic weapons may not be as likely.

## H2: Direct Energy Weapons (DEW)

"***Moscow Signals***". The Cold War drove a persistent arms race between the West and the Eastern Bloc. In 1973, US personnel were briefed on possible microwaves targeting the upper offices of the US Embassy in Moscow [33] [34]. A declassified CIA report validated the threat that microwaves were pointed at the embassy's upper floors and leaked into the building through telephone hardlines [35]. Alternative, hypotheses implied that the Moscow Signal worked to power or activate eavesdropping bugs within the compromised structure of the building [36] [37]. It is suggested that this capability existed since the 1950s. However, in 1976, it is later understood that the mysterious microwaves emissions were attempts to target and jam American Signals Intelligence (SIGINT) capabilities [38] [39]. Depending on the section of the building, an employee could receive varying levels of microwave radiation and over time metal faraday mesh was applied to shield diplomats [38] [40]. One study suggested poor concentration and poor memory were symptoms of exposure to microwaves or *electro-hypersensitivity* [41]. Needless to say, there was significant danger to health of US diplomats then as there is now.

***Historical Precedent***. The US Government cannot forget that the USSR attempted to convince the American public that Acquired Immune Deficiency Syndrome (AIDS) was an American bioweapon [42]. Misinformation campaigns were real and prevalent during the Cold War. It is not beyond reasonable doubt that Direct Energy Weapons have persisted. There is ample evidence of misinformation campaigns by today's adversaries. In the modern day, old adversaries are reapplying old tactics.

Countries may disappear overnight, but tradecraft does not. Although important, these tangent into information warfare are out-of-scope for this paper.

***Present Day Applications***. Naturally, students learn about dangers of chemical warfare in history lessons of World War I and radiation sickness following dropping of atomic bombs at the conclusion of World War II. However, the American People have very little understanding in the ramifications of DEW. These have been seen in Active Denial Technology (ADT) systems or commonly known as *less-lethal* weapons [43] [44]. Less-lethal weapons are commonly employed in riot and crowd control situations as supporting assets in shaping battlefield or urban conditions.

***Electromagnetic Spectrum***. Electromagnetic radiation consists of the entire span of frequencies and wavelengths of energy. Within this spectrum, there consists the visible light spectrum as well as radio waves and high-energy rays [45] [46].

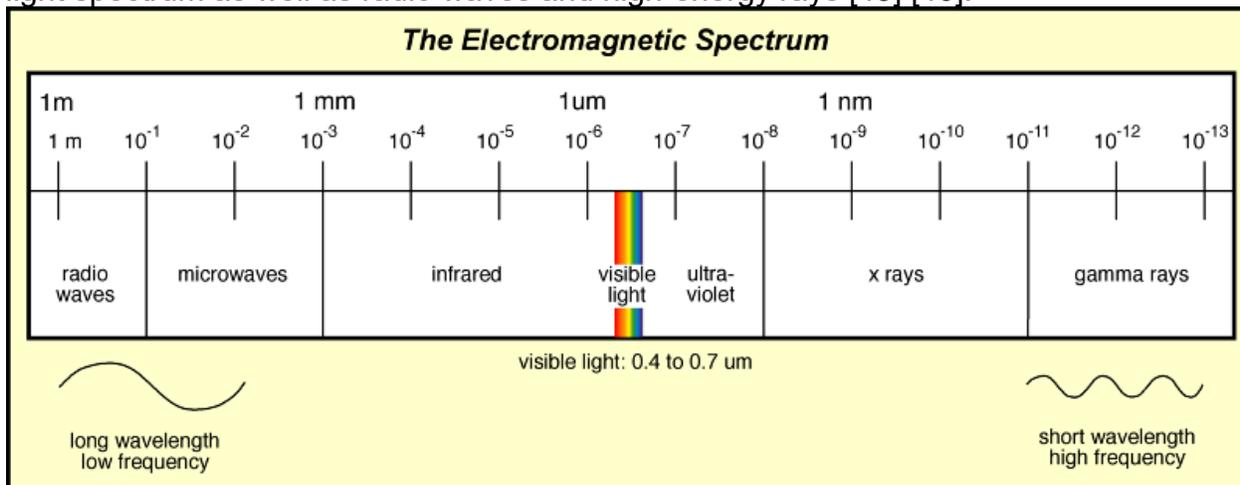

*Figure 2: Electronic Magnetic Spectrum [46] (source: Columbia University)*

***Modern Studies of Effects***. A 2003 study by Motorola Florida Research Laboratories investigated "*human auditory response to pulses of radiofrequency (RF) energy*" [47]. According to the study, people have reported experiencing chirping noises or buzzing and these symptoms have been reported since 1947 during advancements of radar [48]. A 2018 research into USG employees suggested victims suffered from unknown brain injuries and faced unclear sensory manifestations "without an associated history of head trauma" [49] [50] [51]. There are also additional studies which hearing loss was a result of chemotherapy and radiation [52].

***All About the Ear***. Dr. James Giordano of Georgetown University Medical Center suggested that direct energy weapons could create cavitation bubbles within inner ear which contains fluid or blood [53]. Simply put, DEWs could boil the fluid within the vestibular system which can influence a person's balance and could explain the vertigo and nausea. The National Academy of Sciences held similar conclusions in their 2020 assessment of causes for Havana Syndrome [54] [55] [56]. Allegedly, diplomats were found to also have damaged eardrums [57].

***Discussion***. Sound may not always penetrate well-built walls, but the right band of electromagnetic radiation can. For example, Wi-Fi (radio, low frequency) or X-Rays (light, high frequency), these waves can penetrate solid objects and be directed towards other objects, sometimes with high accuracy. Society also forgets that the electromagnetic spectrum operates on duality where this energy exhibits both properties of waves and particles depending on frame of reference. One major flaw with this hypothesis is that there is limited information available on direct energy weapons, if any exist. There are also significant moral and ethical implications for research testing of direct energy on living organisms. Additionally, the Laws of War & Armed Conflict as well as International Conventions of War (Geneva or Hague) have not evolved to consider new age of weapons [58]. The American People must to come to the realization as said by Mark Twain: "*history never repeats itself, but it does often rhyme*".

## H3: Crickets

One hypothesis suggests that cricket chirps are the cause of Havana Syndrome. More specifically, some researchers suggest it was the West Indies Cricket [59]. Although studies have linked the unusual noise to the crickets for some cases, not all diplomats were tested equally with respect to MRIs, blood tests, and toxicological screenings [60] [61]. Crickets can migrate like any other insect or creature, but there is no serious evidence to demonstrate that West Indies crickets have become a seriously-dangerous or invasive species in several very specific regional capitals or globally. Some victims have associated the noise they heard with the crickets. However, there is not enough evidence to demonstrate that crickets have caused the reported concussive injuries.

*Discussion*. Cricket chirps are a great nuisance to those who wish to live a life of peace and absolute quiet. However, it does not explain why only specific individuals from western nations (namely diplomats from both US and Canada) suffered. It's very possible that crickets migrated with trade routes similar to the lethal and highly-aggressive Asian Giant Hornet. Local fauna or crickets could have very well annoyed several diplomats or triggered memories of attacks. The limited pool of victims does not explain the unusual places which the victims were injured such as with Washington, DC, which is not well known for cricket infestations, within a heavily urbanized city. People hear insects and can find them annoying. Insects migrate and can invade new ecosystems. However, crickets do not fully explain victimology, nature of injuries, or oddly specific geographic spread of alleged victims.

## H4: Mass Hysteria

Psychological effects are still possible. With respect to mass hysteria, history has demonstrated the power of suggestion with incidents such as the Salem Witch Trials [1]. Although diplomats sometimes work in austere conditions or in unfriendly countries, victims were not reporting suffering from consistent trauma from people harassing or heckling them. Some researchers have attributed Havana Syndrome to stress from being under constant harassment or surveillance in unfriendly countries [62]. Other scientists have suggested psychosomatic illness [62] [63].

***Discussion***. Mass Hysteria happens in the modern day. A plain example would be the reckless behavior that Americans commit or suffer during midnight openings on Black Fridays. One cannot in good conscience support the mass hysteria or *stress* hypotheses because diplomats have had their brains injured when compared in scans to healthy individuals [64]. Denying the claims of the victims and handwaving their concerns are insults to cumulative decades of faithful service to the American People. Stress and anxiety from seeing others suffer could lead to hypochondria. Hysteria or hypochondriac tendencies are possible. The mind is powerful, but physical injuries are abnormal. Secondary reporting and journals do not suggest these individuals have a history of concussions.

## H5: Chemical Exposure

Some hypotheses suggest neurotoxins or chemical contamination have led to these symptoms. It is not unreasonable to believe this due to reliance on pesticides [64]. Agent Orange caused significant defects in the Vietnamese population [65]. Even today, the global community has yet to fully grasp the ramifications or long-term health effects from consuming water contaminated with microplastics and these microplastics have leached into aquifers and water tables [66]. In recent years, there have been several instances of deliberate heavy metal poisoning or nerve agent attacks [67] [68] [69]. Poisons, toxins, nerve agents, and chemicals are all well-acquainted tactics of total war, espionage, and assassinations. These are hard to trace, but not impossible to identify forensically.

***Discussion***. Poisoning is not completely out of the question. However, in cases of poisoning, there was significant investigative breakthroughs to associate hostile powers or hostile chemicals to an event. With respect to Havana Syndrome, numerous individuals were injured and there were still no leads on injuries? It is unreasonable to suggest that multiple diplomats in various geographically-separate embassies were hurt by the same chemical without anyone noticing deployment of such chemicals deliberately or accidentally. Chemicals are possible as contact or aerosol weapons, but they require an attacker to deploy the chemical onto a surface, area, or victim and be in the general vicinity (in some way, shape, or form). Pesticides may be used by normal hard-working farmers. However, some cases, chemicals causing Havana Syndrome in heavily urbanized regional capitals far away from large swaths of greenery is far-fetched.

## Case Study: Beck

In 1996, Michael Beck was a USG employee serving overseas for brief tours. He reported suffering from injuries consistent with what would be known as Havana Syndrome. In 2014, Beck received an unclassified memorandum (dated 2012) acknowledging intelligence of association between a hostile foreign power and a potentially lethal microwave weapon systems […] during his time of service in a given

Area of Operations [70]. Unfortunately, the memorandum falls short of acknowledging any relationship between Beck's injuries, the alleged microwave system (if it existed in the 1990s), and the alleged adversarial nation in the 1990s.

***Hypothesis of the Case***. Over the course of the Cold War until the modern day, Direct Energy Weapons were very likely continuously researched, developed, and deployed. The injuries and long-term risks associated with harm from DEW are not fully understood because the pool of victims is small, limited transparency on existing adversarial weapons, and limited victim information.

## Activity in the Gray Zone

***Moscow Rules***. The Moscow Rules were an unwritten guide for personnel maneuvering in Moscow during the Cold War. The rules imply that an individual will always be surrounded and monitored by their adversaries [32] [33]. Although the years go by and names change and technology evolves, but adversaries remain constant.

***Current Official Hypothesis***. The current verdict is that there is no evidence to suggest a nation-state attack or weapon. However, it ignores historical precedent of behaviors of nations and the behavior of the evolution of warfare. Weapons will continue to evolve to meet the needs of the battlefield or battlespace. Use of weapons should not be thought of as only contained engagements. If Havana Syndrome represents the results of an attack on diplomats, then it could very well suggest testing of weapons that are nearly impossible to track or defeat.

***Allies Injured***. *Qui bono*? America's adversaries. Old Cold War adversarial nations had everything to gain from disrupting American foreign policy efforts and to prevent mending of relations between the US and Cuba. Historically, Cuba was a USSR ally. There is no such thing as coincidence in geopolitics. There are claims that several Canadian diplomats have been reportedly injured with symptoms similar to Havana Syndrome and of which includes brain injuries [73] [74]. These occurrences are not solely an American issue.

***Temporal Analysis***. Below is merely a table of several documented instances of well-known official travel and several reported instances of Havana Syndrome. Although, there may or may not be a relationship between the two as: *correlation is not causation*.

| *Year* | *Location* | *Geopolitics of the Time* |
|---|---|---|
| March 2016 | Havana, Cuba | President Obama attempted normalization of relations with Cuba [75] [76]. |
| 2017 | Havana, Cuba | Additional reported injuries. |
| 2018 | Guangzhou, China | Diplomats in Guangzhou reported symptoms [77] |
| 2019 | Washington, DC | USG official felt symptoms immediately after leaving a building [78]. |
| Unknown | London, UK | Alleged site of Havana Syndrome attacks [63]. |
| 2021 | Berlin, Germany | Diplomats in Berlin reported symptoms [77] |
| 2021 | Vienna, Austria | Reports of attacks [79]. |

| 2021 | India | First meeting of the Quad, diplomat escorting USG official suffers symptoms [80] [81] |
| 2021 | Hanoi, Vietnam | VP attempts improving relations with Vietnam, new embassy leasing [82]. Two US diplomats evacuated [83]. |

*Geographic Spread*. Some hypotheses do not address the inconsistencies of victim locations and citizenship of victims. Diplomats were affected in Washington (DC), China, Cuba, Vietnam, and India [84]. The conundrum is that Havana Syndrome reportedly only affected a handful of individuals in a select cluster of countries. These incidents are occurring within national or regional capital cities.

*Non-Traditional Weapons & Countermeasures*. US airmen reported midflight injuries by lasers shined by hostile forces [85]. The Second Invasion of Ukraine has given invaluable lessons and insight on what the future of warfare may entail. Camouflage uniforms will eventually attempt to suppress infrared signatures given off by body heat [86]. The US Military prides itself on well-trained and technologically sophisticated capabilities, but what if adversaries have studied what makes us better or better-yet reliant – following Occam's Razor: reliance on technology makes us complacent. It is not unreasonable to believe that adversaries are investing in alternative capabilities for which the US has not thought of or invested in. Our reliance on technology such as satellites are equally. *What does this have to do with Havana Syndrome*? Non-traditional weapons will create conditions for asymmetric warfare to challenge American military superiority in kinetic warfare [87]. Simply put, *why fight fairly when there is no incentive to*?

*Discussion*. While military engagements are projections of hard power, diplomacy and economics are implementations of soft power. These types of power are the applications of influence. Now, war is a means to an end when diplomacy fails and is a geopolitical endeavor. Along the spectrum of peace and war, there exists the transient area, commonly referred to as the *gray zone,* where applications or actions are limit-tests and calculated or optimized engagements leveraged to harass, deter, disrupt, defeat, or destroy oppositional forces without being enough to justify for kinetic warfare [88]. At CYBERWARCON 2023, Emerson T. Brooking of Atlantic Council suggested that the difference between information operations and cyber operations comes from the temporal nature of the activity; information operations focuses on long-term influence of narrative and shaping human-thinking where cyber operations are short-term and ephemeral in nature: steal information and scram [89]. In addition, society is slowly beginning to see the third intersection of digital warfare – information, cyber, and now electronic warfare.

## Analysis of Competing Hypotheses

*Discussion*. It is a personal opinion that these injuries appear consistent with expected results of Direct Energy Weapons. Although all hypotheses need to be considered, it is impractical to believe this may be the result of chemical contamination based on the type of injuries. Additionally, the hypothesis that crickets (and disease) caused such activity does not fully account the victimology and clustered geographic spread of alleged attacks. If this were the fault of crickets, then numerous countries would be facing crises of invasive species and the victim pool would be far greater.

From paranoia, it can be argued that many targets were not high-level officials, but rather lower-echelon or mid-tier personnel. This would keep engagements persistent, but plausibly deniable without incurring wrath of victim nations or warranting suspicion. This feeds into adversarial adaptation of their own *Gray Zone* (or Hybrid Warfare) doctrine. Countries may not wish direct military confrontation with the US, given US superior capabilities, logistics, and training. However, low-intensity engagements and investments into non-traditional weapons would be a possible solution to degrading or minimizing likelihood of eliciting a US military response and thus avoiding direct-action conflicts.

## Recommendations

The concern for safety US diplomats and those injured while serving abroad in diplomatic missions cannot be ignored. Based on this research, Congress is implored to reopen the investigation into Havana Syndrome. This is not to disparage the years of work by the fine analysts of other researchers. This is a whitepaper based on concerns that Americans may have been injured by unnatural causes. There is no loss of value from re-examining these medical ailments and records. Similar to persistent engagement in cyber operations, the United States needs to engage and shame our adversaries for resorting to uncouth tactics to harm diplomats in this global game of thrones.

The following is recommended:

1. Congress must be informed of adversarial scientific advances in non-traditional kinetic weaponry.

2. Congress should reopen the investigation into Havana Syndrome, given that the current hypotheses do not sufficiently answer to the reported injuries.

3. Congress should invest and allocate resources into researching countermeasures and defenses against non-traditional weapons (ie: electromagnetic or sonic).

4. Similar to cyber deterrence through *name-and-shame*, the US should be prepared to do the same, that is if a geopolitical adversary has been discovered to be leveraging a supposed non-traditional weapon system against officials, diplomats, or US servicemembers.

# Conclusion

It is a personal assessment based on publicly available information that Havana Syndrome could be the result of a direct-energy weapon that possibly target the vestibule system within the inner ear. This would explain injuries to eardrums. The electromagnetic radiation would explain brain injuries to white matter and concussion-like symptoms and disorientation. The world is approaching on a potential global conflict. It is in the best interest to keep the US Congress well-informed of possible adversarial capabilities. As one place themselves in the mindset of an adversary. Havana Syndrome could be a progression of DEW research and vestiges from the Cold War. The United States cannot ignore the fact that our enemies also have their own budgets, their own agendas, and their own defense research initiatives. One cannot help but refer back to a memorable line in *Indiana Jones and the Kingdom of the Crystal Skull*. Befitting of the events and paraphrasing Cate Blanchett's character: "*…all of you, from the inside. […] and the best part? You won't even know it's happening.*" Otherwise, this is merely baseless conjecture and finding the truth to anything is quixotic.

# Future Work

Given that medical information is heavily regulated, continuation of this private research will not be feasible alone. However, this research whitepaper could be carried forward by the USG, if evidence surrounding Havana Syndrome re-examined.

# Conflict of Interest

The author is not a medical provider in any capacity and is a member of the New Hampshire Army National Guard (NH ARNG). The research, discussions, analyses, and conclusions are the author's own and not representative of any view of the Department of Defense or US Government. All research was conducted with publicly-available information. Given a state of geopolitical uncertainty, conclusions may be heavily biased.

# Funding

This research was conducted without grants or funding. This was whitepaper was written as a personal project in pursuit completion of the PhD in Cyber Defense Program at Dakota State University (CAE-CD).